\documentclass[12pt]{article}
\usepackage{a4wide}
\usepackage{amstext,amsfonts,amsmath}
\renewcommand{\appendix}{%
   \setcounter{section}{0}
   \setcounter{subsection}{0}
   \renewcommand{\thesection}{Appendix}%
   \renewcommand{\thesubsection}{A\arabic{subsection}}%
   \renewcommand{\theequation}{A.\arabic{equation}}%
   \setcounter{equation}{0}%
}

\begin{document}

\begin{flushright}
{}FTUV--04/0601 \quad IFIC--04/23
\end{flushright}

\begin{center}

\begin{Large}
{\bf On the underlying gauge group  structure of $D=11$
supergravity}
\end{Large} \vskip 1cm

\begin{large}
I.A. Bandos$^{a,c,1}$, J.A.~de~Azc\'arraga$^{a,1}$, J.M.
Izquierdo$^{b,1}$, M.~Pic\'on$^{a,1}$ and
O.~Varela$^{a,}$\footnote{bandos@ific.uv.es,
j.a.de.azcarraga@ific.uv.es, izquierd@fta.uva.es,
moises.picon@ific.uv.es, \\ oscar.varela@ific.uv.es} \end{large}
\vspace*{0.6cm}

\begin{it}
$^a$ Departamento de F\'{\i}sica Te\'orica, Facultad de
{}F\'{\i}sica, Universidad de Valencia\\ and IFIC,
Centro Mixto Universidad de Valencia--CSIC, \\
E--46100 Burjassot (Valencia), Spain \\
$^b$ Departamento de F\'{\i}sica Te\'orica,
Universidad de Valladolid, \\
E--47011 Valladolid, Spain \\
$^c$ Institute for Theoretical Physics, NSC ``Kharkov Institute of
Physics  and Technology'',
 UA61108, Kharkov, Ukraine \\[0.4cm]
\end{it}
\end{center}

\begin{abstract}
The underlying gauge group structure of $D=11$ supergravity is
revisited. It may be described by a one-parametric family of Lie
supergroups ${\tilde{\Sigma}(s) \times\!\!\!\!\!\!\supset
SO(1,10)}$, $s \neq 0$. The family of superalgebras
$\tilde{\mathfrak{E}}(s)$ associated to $\tilde{\Sigma}(s)$ is
given by a family of extensions of the M-algebra $\{P_a, Q_\alpha,
Z_{ab}, Z_{a_1\ldots a_5} \}$ by an additional fermionic central
charge $Q^\prime_\alpha$. The Chevalley-Eilenberg four-cocycle
$\omega_4 \sim \Pi^\alpha \wedge \Pi^\beta \wedge \Pi^a \wedge
\Pi^b \Gamma_{ab}{}_{\alpha\beta}$ on the standard $D=11$
supersymmetry algebra may be trivialized on
$\tilde{\mathfrak{E}}(s)$, and this implies that the three-form
field $A_3$ of $D=11$ supergravity may be expressed as a composite
of the $\tilde{\Sigma}(s)$ one-form gauge fields $e^a$,
$\psi^\alpha$, $B^{ab}$, $B^{a_1\ldots a_5}$ and $\eta^\alpha$.
Two superalgebras of $\tilde{\mathfrak{E}}(s)$ recover the two
earlier D'Auria and Fr\'e decompositions of $A_3$. Another member
of $\tilde{\mathfrak{E}}(s)$ allows for a simpler composite
structure for $A_3$ that does not involve the $B^{a_1\ldots a_5}$
field. $\tilde{\Sigma}(s)$ is a deformation of
$\tilde{\Sigma}(0)$, which is singularized by having an enhanced
$Sp(32)$ (rather than just $SO(1,10)$) automorphism symmetry and
by being an expansion of $OSp(1|32)$.
\end{abstract}

\section{Introduction}

M-theory (see \cite{M-theory}) emerged at the time of the second
superstring revolution in the mid nineties. In contrast with other
theories like the standard model, QCD or general relativity,
M-theory is at present not based on a definite Lagrangian or on an
S-matrix description; rather, it is characterized by its different
perturbative and low energy limits (string models and
supergravities) and by dualities \cite{Duality} among them. Such
dualities, including those relating apparently different models,
are believed to be symmetries of M-theory; the full set of
M-theory symmetries\footnote{Several groups may play a role, as
the rank 11 Kac-Moody $E_{11}$ group \cite{WestE11} or $OSp(1|64)$
\cite{Bars98,West00} and its subgroup $GL(32)$ \cite{BWest00,GGT}.
This group is the automorphism group of the M-algebra $\{Q_\alpha
, Q_\beta \} = P_{\alpha \beta}$; it is also a manifest symmetry
of the actions \cite{BL98,BPS04} for BPS preons \cite{BPS01}, the
hypothetical constituents of M-theory. Clearly, in $D=11$
supergravity one might see only a fraction of the M-theory
symmetries. As it was noticed recently
\cite{DL03,Hull03} (see also \cite{BPS04}), a suggestive analysis
of partially supersymmetric $D=11$ supergravity solutions can be
carried out in terms of generalized connections with holonomy group
$SL(32)$. The case for a $OSp(1|32)\otimes OSp(1|32)$ gauge
symmetry in a Chern-Simons context was presented in
\cite{Zanelli,Horava,Nastase}.} should include these dualities as
well as the symmetries of the different superstring and
supergravity limits.

In this letter we are interested in the underlying gauge symmetry
of $D=11$ supergravity as a way of understanding the symmetry
structure of M-theory. The problem of the hidden or underlying
geometry of $D=11$ supergravity was raised already in the
pioneering paper by Cremmer-Julia-Scherk (CJS) \cite{CJS} (see
also \cite{BAMcDo83,Kallosh84}), where the possible relevance of
$OSp(1|32)$ was suggested. It was specially considered by D'Auria
and Fr\'e \cite{D'A+F}, where the search for the local supergroup
of $D=11$ supergravity was formulated as a search for a composite
structure of its three-form $A_3$. Indeed, while the graviton and
gravitino are given by one-form fields $e^a=dx^\mu e_\mu^a(x)$,
$\psi^\alpha= dx^\mu \psi^\alpha_\mu(x)$ and can be considered,
together with the spin connection
$\omega^{ab}=dx^\mu\omega_\mu^{ab}(x)$,  as gauge fields for the
standard superPoincar\'e group \cite{DoMa77}, the
$A_{\mu_1\mu_2\mu_3}(x)$ abelian gauge field is not associated
with a symmetry generator and it rather corresponds to a
three-form $A_3$. However, one may ask whether it is possible to
introduce a set of additional one-form fields such that they,
together with $e^a$ and $\psi^\alpha$, can be used to express
$A_3$ in terms of products of one-forms. If so, the `old' and
`new' one-form fields may be considered as gauge fields of a
larger supergroup, and all the CJS supergravity fields can then be
treated as gauge fields, with $A_3$ expressed in terms of them.
This is what is meant by the underlying gauge group structure of
$D=11$ supergravity: it is hidden when the standard $D=11$
supergravity multiplet is considered, and manifest when $A_3$
becomes a composite of the one-form gauge fields associated with
the extended group. The solution to this problem is equivalent (see
Sec.~\ref{trivializarion}) to trivializing a
standard $D=11$ supersymmetry algebra four-cocycle (related to
$dA_3$) on an enlarged superalgebra.

Two superalgebras with a set of $528$ bosonic and $32+32=64$
fermionic generators
\begin{equation} \label{generators}
 \, P_a  \, , \, Q_\alpha  \, , \,  Z_{a_1a_2}  \, ,
\, Z_{a_1 \ldots a_5}  \, , \, Q^\prime_\alpha \; ,
\end{equation}
including the M-algebra \cite{M-alg} ones plus a central fermionic
generator $Q^\prime_\alpha$, were found in \cite{D'A+F} to allow
for a decomposition of $A_3$. Both superalgebras are clearly
larger than $osp(1|32)$, but an analysis \cite{CFGPN} of its
possible relation with $osp(1|64)$ and $su(1|32)$ (by an
\.In\"on\"u-Wigner contraction) gave a negative answer. The two
D'Auria-Fr\'e superalgebras are particular elements (namely,
$\tilde{\mathfrak{E}}(3/2)$ and $\tilde{\mathfrak{E}}(-1)$) of a
one-parametric family of superalgebras $\tilde{\mathfrak{E}}(s)$
characterized by specific structure constants, the meaning of
which has been unclear until present.

In fact, the first message of this letter is that the underlying
gauge supergroup structure of the $D=11$ supergravity can be
described by any representative of a {\it one-parametric family
of supergroups ${\tilde{\Sigma}(s) \times\!\!\!\!\!\!\supset
SO(1,10)}$ for $s\not=0$}, and that these are nontrivial
($s\not=0$)  deformations of ${\tilde{\Sigma}(0)
\times\!\!\!\!\!\!\supset SO(1,10)} \subset {\tilde{\Sigma}(0)
\times\!\!\!\!\!\!\supset Sp(32)} $, where
${\times\!\!\!\!\!\!\supset}$ means semidirect product. The second
point is the relation of the underlying gauge supergroups with
$OSp(1|32)$.  Recently, a new method for obtaining Lie algebras
from a given one has been proposed in \cite{H01} and developed in
\cite{AIPV02}. The relevant feature of this procedure, the {\it
expansion method} \cite{AIPV02} is that, although it includes the
\.In\"on\"u-Wigner contraction as a particular case, it is not a
dimension preserving process in general, and leads to
(super)algebras of higher dimension than the (super)algebras that
are expanded. We show that ${\tilde{\Sigma}(0)
\times\!\!\!\!\!\!\supset SO(1,10)}$ may be obtained from $OSp(1|32)$
by an expansion: ${\tilde{\Sigma}(0) \times\!\!\!\!\!\!\supset
SO(1,10)} \approx OSp(1|32)(2,3,2)$ (see Appendix). The $SO(1,10)$
automorphism group of $\tilde{\Sigma}(s)$ is enhanced to $Sp(32)$
for $\tilde{\Sigma}(0)$.  It is also seen that ${\tilde{\Sigma}(0)
\times\!\!\!\!\!\!\supset Sp(32)}$ is the expansion
$OSp(1|32)(2,3)$.

\section{Trivialization of a
Chevalley-Eilenberg four-co\-cy\-cle and composite nature of the
$A_3$ field}

\label{trivializarion}

Supergravity is a theory of local supersymmetry. The graviton
$e_\mu^a(x)$ and the gravitino $\psi^\alpha_\mu(x)$ can be
considered as gauge fields associated with the standard
supertranslations algebra $\mathfrak{E}$ ($\,\equiv
\mathfrak{E}^{(D|n)}$ in general, $\mathfrak{E}^{(11|32)}$ for
$D=11$),
\begin{eqnarray}
\label{QQ=P} {} \{ Q_\alpha , Q_\beta \} = \Gamma^a_{\alpha\beta}
P_a \; , \qquad [P_a, Q_\alpha]=0 \; , \quad [P_a, P_b]=0 \; .
\end{eqnarray}
The supergravity one-forms $e^a$, $\psi^\alpha$ and $\omega^{ab}$
(spin connection) generate a free differential algebra
(FDA)\footnote{In essence, a FDA (introduced in this context in
\cite{D'A+F} as a {\it Cartan integrable system}) is an exterior
algebra of forms, with constant coefficients, that is closed under
the exterior derivative $d$; see \cite{Su77,D'A+F,Ni83}.} defined
by the expressions for the FDA curvatures
\begin{eqnarray}\label{Ta=}
{\mathbf R}^a &:=&  de^a -e^b\wedge \omega_b{}^a  + i\psi^{\alpha}
\wedge \psi^{\beta} \Gamma^a_{\alpha\beta} = T^a  + i\psi^{\alpha}
\wedge \psi^{\beta}
\Gamma^a_{\alpha\beta} \; , \\
\label{Tal=} {\mathbf R}^\alpha  &:= &  d\psi^\alpha - \psi^\beta
\wedge \omega_\beta{}^\alpha  \quad \left(
\omega_\alpha{}^\beta={1\over 4} \omega^{ab} \Gamma_{ab\;
\alpha}{}^\beta \right) \; ,
\\
\label{Rab=} \mathbf{R}^{ab} &:= &  d \omega^{ab}  - \omega^{ac}
\wedge \omega_c{}^{b}  \; ,
\end{eqnarray}
where $T^a:=De^a=de^a -e^b\wedge \omega_b{}^a$ is the torsion and
$\mathbf{R}^{ab}$ coincides with the Riemann curvature, and by the
requirement that they satisfy the Bianchi identities that
constitute the selfconsistency or integrability conditions for
Eqs.~(\ref{Ta=})--(\ref{Rab=}). When all curvatures are set to
zero, ${\mathbf R}^a=0$, ${\mathbf R}^\alpha =0$, ${\mathbf
R}^{ab} =0$,  Eqs.~(\ref{Ta=}) and (\ref{Tal=}) reduce, if we
remove the Lorentz $\omega^{ab}$ part, to the Maurer-Cartan (MC)
equations for $\mathfrak{E}$,
\begin{eqnarray}\label{MCTa=}
de^a = - i\psi^{\alpha} \wedge \psi^{\beta} \Gamma^a_{\alpha\beta}
\; , \qquad d\psi^\alpha =0 \; .
\end{eqnarray}

One easily solves (\ref{MCTa=}) by
\begin{eqnarray}\label{ea=Pia}
e^a = \Pi^a := dx^a - i d\theta^\alpha  \Gamma^a_{\alpha\beta}
\theta^\beta \; , \qquad
\psi^{\alpha}=\Pi^{\alpha}:=d\theta^\alpha \; ,
\end{eqnarray}
where $\Pi^a$, $\Pi^\alpha$ are the MC forms for the
supertranslation algebra.  Considered as forms on rigid superspace
($\Sigma^{(D|n)}$ in general), one identifies $x^a$ and
$\theta^\alpha$ with the coordinates $Z^M=(x^a,\theta^\alpha)$ of
this superspace\footnote{ {\it Rigid} superspace is the group
manifold of the supertranslations group $\Sigma^{(D|n)}$. We shall
use the same symbol $\Sigma^{(D|n)}$, $\tilde{\Sigma}$, to denote
both the supergroups and their manifolds.}. When $e^a$ and
$\psi^\alpha$ are forms on spacetime, $x^a$ are still spacetime
coordinates while $\theta^\alpha$ are Grassmann functions,
$\theta^\alpha=\theta^\alpha(x)$, the Volkov-Akulov Goldstone
fermions \cite{VA72}. For one-forms defined on curved standard
superspace, $e^a=dZ^ME_M^a(Z)$, $\psi^\alpha=dZ^ME_M^\alpha(Z)$,
$\omega^{ab}(Z)= dZ^M\omega_M^{ab}(Z)$ the FDA (\ref{Ta=}),
(\ref{Tal=}), (\ref{Rab=}) with nonvanishing ${\mathbf R}^\alpha$
and ${\mathbf R}^{ab}=R^{ab}$ but vanishing ${\mathbf R}^a=0$
gives a set of superspace supergravity constraints (which are
kinematical or {\it off-shell} for $D=4$, $N=1$ and {\it
on-shell}, {\it i.e.} containing equations of motion among their
consequences, for higher $D$ including $D=11$ \cite{Howe97}).
However, the FDA makes also sense for forms on spacetime, where
$e^a=dx^\mu e_\mu^a(x)$ and $\psi^\alpha=dx^\mu
\psi_\mu^\alpha(x)$ are the gauge fields for the supertranslations
group.

{}For $D=11$ supergravity, however, the above FDA description is
incomplete since the CJS supergravity supermultiplet includes, in
addition to $e_\mu^a(x)$ and $\psi^\alpha(x)$, the antisymmetric
tensor field $A_{\mu\nu\rho}(x)$ associated with the three-form
$A_3$. The FDA (\ref{Ta=}), (\ref{Tal=}), (\ref{Rab=}) has to be
completed by the definition of the four-form field strength
\cite{D'A+F}
\begin{eqnarray}\label{R4=} \mathbf{R}_4 &:=&
dA_3 + {1\over 4} \psi^\alpha \wedge \psi^\beta \wedge e^a \wedge
e^b \Gamma_{ab}{}_{\alpha\beta} \; .
\end{eqnarray}
Note that, considering the FDA (\ref{Ta=}), (\ref{Tal=}),
(\ref{Rab=}), (\ref{R4=}) on the $D=11$ superspace and setting
$\mathbf{R}^a=0$ and $\mathbf{R}_4=F_4:= 1/4! e^{a_4}\wedge \ldots
\wedge e^{a_1}F_{a_1\ldots a_4}$ one arrives at the original
on-shell $D=11$ superspace supergravity constraints
\cite{CremmerFerrara80,BrinkHowe80}. But, and in contrast with the
$D=4$ case, the above FDA for vanishing curvatures cannot be
associated with the MC equations of a {\it Lie} superalgebra due
to the presence of the {\it three}-form $A_3$. However, on rigid
superspace $\Sigma^{(11|32)}$ (the group manifold of the $D=11$
supertranslations group), where one also sets  $\mathbf{R}_4=0$ by
consistency, the bosonic four-form
\begin{eqnarray}\label{a4=}
a_4= - {1\over 4} \psi^\alpha \wedge \psi^\beta \wedge e^a \wedge
e^b \Gamma_{ab}{}_{\alpha\beta} \;
\end{eqnarray}
becomes a Chevalley-Eilenberg (CE) \cite{CE,book} Lie algebra
cohomology four-cocycle on $\mathfrak{E}$,
\begin{eqnarray}\label{a4=0}
\omega_4 (x^a, \theta^\alpha) &=& - {1\over 4} \Pi^\alpha \wedge
\Pi^\beta \wedge \Pi^a \wedge \Pi^b
\Gamma_{ab}{}_{\alpha\beta}=d\omega_3 (x^a, \theta^\alpha)
\end{eqnarray}
since $\omega_4$ is invariant and closed. The
$\mathfrak{E}^{(11|32)}$ four-cocycle $\omega_4$ is, furthermore,
a non-trivial CE one, since the above three-form
$\omega_3=\omega_3(x^a,\theta^\alpha)$ cannot be expressed in
terms of the invariant MC forms of $\mathfrak{E}^{(11|32)}$. Now,
we may ask whether there exists an {\it extended} Lie
superalgebra, generically denoted $\tilde{\mathfrak{E}}$, with MC
forms on its associated extended superspace $\tilde{\Sigma}$, on
which the CE four-cocycle $\omega_4$ becomes trivial. In
this way, the problem of writing the original $A_3$ field in terms
of one-form fields becomes purely geometrical: it is equivalent to
looking, in the spirit of the fields/superspace variables
democracy of \cite{JdA00}, for an {\it enlarged} supergroup
manifold $\tilde{\Sigma}$ on which one can find a new three-form
$\tilde{\omega}_3$ (corresponding to $A_3$) written in terms of
products of $\tilde{\mathfrak{E}}$ MC forms on $\tilde{\Sigma}$
(corresponding to one-form gauge fields) that depend on the
coordinates $\tilde{Z}$ of $\tilde{\Sigma}$. That such a form
$\tilde{\omega}_3(\tilde{Z})$ should exist here is also not
surprising if we recall that the $(p+2)$- CE cocycles on
$\mathfrak{E}$ that characterize \cite{AzTo89} the WZ terms of the
super-$p$-brane actions and their associated FDA's, can also be
trivialized on larger superalgebras $\tilde{\mathfrak{E}}$
\cite{BeSez95,JdA00} associated to extended superspaces
$\tilde{\Sigma}$, and that the pull-back of
$\tilde{\omega}_3(\tilde{Z})$ to the supermembrane worldvolume
defines an invariant WZ term.

The MC equations of the larger Lie superalgebra
$\tilde{\mathfrak{E}}^{(11|32)}$ trivializing $\omega_4$ can be
`softened' by adding the appropriate curvatures. Considering the
resulting FDA for the `soft' forms over eleven-dimensional
spacetime, one arrives at a theory of $D=11$ supergravity in which
$A_3$ is a {\it composite}, not elementary, field. Its FDA
curvature, $R_4$ in Eq.~(\ref{R4=}), is then expressed through the
curvatures of the old and new one-form gauge fields.

\section{A family of extended superalgebras $\tilde{\mathfrak
E}(s)$ allowing for a trivialization of the CE four-co\-cy\-cle
$\omega_4$}

It was found in \cite{D'A+F} that it was possible to write the
three-form $A_3$ of the $D=11$ supergravity FDA (\ref{Ta=}),
(\ref{Tal=}), (\ref{Rab=}), (\ref{R4=}) in terms of one-forms, at
the prize of introducing two new bosonic one-forms, $B^{a_1 a_2}$,
$B^{a_1 \ldots a_5}$, and one new fermionic one-form
$\eta^\alpha$, obeying the FDA  equations
\begin{eqnarray}
\label{cBab} {\cal B}_2^{a_1a_2} &=& DB^{a_1a_2} + \psi^\alpha
\wedge \psi^\beta \, \Gamma^{a_1a_2}_{\alpha\beta} \; , \qquad
\\
\label{cB15} {\cal B}_2^{a_1\ldots a_5} &=& DB^{a_1\ldots a_5} + i
\psi^\alpha  \wedge \psi^\beta \,
\Gamma^{a_1\ldots a_5}_{\alpha\beta} \; , \qquad \\
 \label{cBal}
{\cal B}_2^\alpha  &=& D\eta^{\alpha} - i \, \delta \;  e^a \wedge
\psi^\beta \Gamma_{a\, \beta}{}^\alpha - \gamma_1 \, B^{ab} \wedge
\psi^\beta \Gamma_{ab\, \beta}{}^\alpha - i \, \gamma_2 \,
B^{a_1\ldots a_5} \wedge \psi^\beta \Gamma_{a_1\ldots a_5
\beta}{}^\alpha \;, \qquad
\end{eqnarray}
{}for two sets of specific values of the parameters, namely
\begin{equation}\label{parameters}
\delta=5\gamma_1 \;,\quad \gamma_2=\gamma_1/(2\cdot4!) \quad
(\gamma_1\neq 0) \qquad \mathrm{and} \qquad \delta=0 \;,\quad
\gamma_2=\gamma_1 /(3\cdot4!) \quad (\gamma_1\neq 0) \;.
\end{equation}

{}For vanishing curvatures and spin connection, $\omega^{ab}=0$,
Eqs.~(\ref{cBab})--(\ref{cBal}) read
\begin{eqnarray}
\label{dB2=} dB^{a_1a_2} &=& - \psi^\alpha \wedge \psi^\beta \,
\Gamma^{a_1a_2}_{\alpha\beta} \; , \qquad \\
\label{dB5=}  dB^{a_1\ldots a_5} &=& - i \psi^\alpha \wedge
\psi^\beta \, \Gamma^{a_1\ldots a_5}_{\alpha\beta} \; ,  \\
 \label{deta=}
d\eta^{\alpha} &=& \psi^\beta \wedge \left(- i \, \delta \, e^a
\Gamma_{a\, \beta}{}^\alpha - \gamma_1 \, B^{ab} \Gamma_{ab\,
\beta}{}^\alpha - i \, \gamma_2 \, B^{a_1\ldots a_5}
\Gamma_{a_1\ldots a_5 \beta}{}^\alpha \right) \; . \qquad
\end{eqnarray}
Eqs.~(\ref{MCTa=}) together with Eqs.~(\ref{dB2=})--(\ref{deta=})
provide the MC equations for the superalgebra
\begin{equation}\label{susyalg}
\{Q_\alpha,Q_\beta\}=\Gamma^a_{\alpha\beta} P_a +
i\Gamma^{a_1a_2}_{\alpha\beta} Z_{a_1a_2} + \Gamma^{a_1\ldots
a_5}_{\alpha\beta} Z_{a_1\ldots a_5} \; ,
\end{equation}
\begin{eqnarray}\label{[Z,Q]}
[ P_a , Q_\alpha ] &=& \delta \;  \Gamma_{a \; \alpha}{}^\beta
Q^\prime_\beta \; , \quad \nonumber
\\ {} [ Z_{a_1a_2} , Q_\alpha ]=i\gamma_1
\Gamma_{a_1a_2 \; \alpha}{}^\beta Q^\prime_\beta \; , &&  \quad [
Z_{a_1 \ldots a_5} , Q_\alpha ]=\gamma_2 \Gamma_{a_1 \ldots a_5 \;
\alpha}{}^\beta Q^\prime_\beta \;   .
\end{eqnarray}
Actually, Eqs.~(\ref{dB2=})--(\ref{deta=}) and
(\ref{susyalg})--(\ref{[Z,Q]}) are not restricted to the cases of
Eq.~(\ref{parameters}); it is sufficient that
\begin{equation}
\label{idg} \delta + 10 \gamma_1- 6! \gamma_2=0 \; ,
\end{equation}
as required by the Jacobi identities \cite{D'A+F}.

One parameter ($\gamma_1$ if nonvanishing, $\delta$ otherwise) can
be removed by rescaling the new fermionic generator
$Q^\prime_\alpha$ and it is thus inessential. Hence
Eqs.~(\ref{susyalg})--(\ref{idg}) describe, effectively, a
one-parameter family of Lie superalgebras that may be denoted
$\tilde {\mathfrak E}(s)$ by using a parameter $s$ given
by\footnote{The case $\gamma_1 \rightarrow 0$, $s \rightarrow
\infty$, may be included with $\gamma_1 s \rightarrow \delta/2
\neq 0$. The corresponding algebra can be denoted
$\tilde{\mathfrak{E}}(\infty)$.}
\begin{eqnarray}
\label{s-def} s:= {\delta \over 2\gamma_1} - 1 \;,\; \gamma_1\neq0
\quad\; \qquad & \Rightarrow & \qquad \left\{
{\setlength\arraycolsep{2pt}\begin{array}{lll} \delta
&=& 2\gamma_1(s+1) \, , \\
\gamma_2 &=& 2\gamma_1({s \over 6!} + {1 \over 5!}) \; .
\end{array}} \right.    \;
\end{eqnarray}
In terms of $s$, Eq.~(\ref{[Z,Q]}) reads:
\begin{eqnarray}\label{Sigma(s)}
[ P_a , Q_\alpha ] &=& 2\gamma_1(s+1) \;  \Gamma_{a \;
\alpha}{}^\beta Q^\prime_\beta \; , \quad \nonumber
\\ {} [ Z_{a_1a_2} , Q_\alpha ]=i\gamma_1
\Gamma_{a_1a_2 \; \alpha}{}^\beta Q^\prime_\beta \; , &&  \quad [
Z_{a_1 \ldots a_5} , Q_\alpha ]=2\gamma_1({s \over 6!} + {1 \over
5!}) \Gamma_{a_1 \ldots a_5 \; \alpha}{}^\beta Q^\prime_\beta \; ,
\end{eqnarray}
and the MC equations for $\tilde {\mathfrak E}(s)$ are given by
Eqs.~(\ref{MCTa=}), (\ref{dB2=}), (\ref{dB5=}) and
 {\setlength\arraycolsep{1pt}
\begin{eqnarray}\label{MCSigma(s)}
d\eta^{\alpha} &=& -2\gamma_1\psi^\beta \wedge \left( i \, (s+1)
\, e^a \Gamma_{a\, \beta}{}^\alpha + \frac12 \, B^{ab}
\Gamma_{ab\, \beta}{}^\alpha + i \, \left({s \over 6!} + {1 \over
5!}\right) \, B^{a_1\ldots a_5} \Gamma_{a_1\ldots a_5
\beta}{}^\alpha \right)  . \quad
\end{eqnarray}}

The $\tilde {\mathfrak E}(s)$ family includes the two
superalgebras \cite{D'A+F} of Eq.~(\ref{parameters}); they
correspond to $\tilde {\mathfrak E}(3/2)$ and $\tilde {\mathfrak
E}(-1)$. We show below, however, that the CE trivialization of
$\omega_4$ is possible for all the $\tilde{\mathfrak{E}}(s)$
algebras but for $\tilde{\mathfrak{E}}(0)$ {\it i.e.}, for all but
one values of the constants $\delta/\gamma_1 , \gamma_2/\gamma_1$
obeying Eq. (\ref{idg}). For these, there exists a
$\tilde{\omega}_3$, $d\tilde{\omega}_3=\omega_4$, that may be
written in terms of the $\tilde {\mathfrak E}(s)$ MC one-forms
defined on the enlarged superspace group manifold
$\tilde{\Sigma}(s)$, $s \neq 0$. Such a trivialization will lead
to a composite structure of the 3-form field $A_3$ in terms of
one-form gauge fields of $\tilde{\Sigma}(s)$.

The $\tilde {\mathfrak E}(0)$ superalgebra constitutes a special
case. It can be written as
\begin{eqnarray}
\label{osp-exp} {} \{Q_\alpha,Q_\beta\}= P_{\alpha\beta} \; ,
\quad [P_{\alpha\beta} , Q_{\gamma}] = 64 \; \gamma_1 \; C_{\gamma
(\alpha } Q^\prime_{\beta )} \; ,
\end{eqnarray}
which follows indeed from Eqs. (\ref{Sigma(s)}),
(\ref{MCSigma(s)}) (cf.~(\ref{susyalg})) because for $s=0$ one can
use the Fierz identity
\begin{eqnarray}
 \label{II=GG}
\delta_{(\alpha}{}^{\gamma} \delta_{\beta)}{}^{\delta} &=& {1\over
32} \Gamma^a_{\alpha\beta} \Gamma_a^{\gamma\delta} - {1\over 64}
\Gamma^{a_1a_2}{}_{\alpha\beta}\Gamma_{a_1a_2}{}^{\gamma\delta} +
{1\over 32 \cdot 5!} \Gamma^{a_1\ldots a_5}{}_{\alpha\beta}
\Gamma_{a_1\ldots a_5}{}^{\gamma\delta}\quad .
\end{eqnarray}
Similarly, it is possible to collect the bosonic one-forms $e^a$,
$B^{a_1a_2}$, $B^{a_1\cdots a_5}$ in Eqs. (\ref{MCTa=}),
(\ref{dB2=}), (\ref{dB5=}) and (\ref{MCSigma(s)}) with $s=0$ in a
symmetric spin-tensor one-form ${\cal E}^{\alpha\beta}$,
\begin{eqnarray}
 \label{cEff=def}
{\cal E}^{\alpha\beta}&=& {1\over 32}\left(e^a
\Gamma_{a}^{\alpha\beta} - {i\over 2}
B^{a_1a_2}\Gamma_{a_1a_2}{}^{\alpha\beta}  + {1\over 5!}
B^{a_1\ldots a_5} \Gamma_{a_1\ldots a_5}{}^{\alpha\beta} \right)\;
,
\end{eqnarray}
that allows us to write the MC equations of
$\tilde{\mathfrak{E}}(0)$ in compact form as
\begin{eqnarray}\label{compacts0}
d{\cal E}^{\alpha\beta}=- i \psi^\alpha  \wedge \psi^\beta \; ,
\qquad d\psi^\alpha=0  \; , \qquad d\eta^{\alpha}=-64i
\gamma_1 \,
\psi^\beta \wedge {\cal E}_\beta{}^\alpha \; ;
\end{eqnarray}
Eqs.~(\ref{osp-exp}) or (\ref{compacts0}) exhibit the $Sp(32)$
automorphism symmetry of $\tilde{\mathfrak E}(0)$.

All the $\tilde {\mathfrak E}(s)$ superalgebras, $s\not=0$, can be
considered as deformations of $\tilde {\mathfrak E}(0)$.
Furthermore, the $\tilde {\mathfrak E}(0)$ superalgebra is singled
out because its full automorphism group is $Sp(32)$ while,
$\forall s\neq0$, $\tilde {\mathfrak E}(s)$ has the smaller
$SO(1,10)$ group of automorphisms. Hence, the generalizations of
the superPoincar\'e group for the $s\neq 0$ and $s=0$ cases are
the semidirect products ${\tilde{\Sigma}(s)
\times\!\!\!\!\!\!\supset SO(1,10)}$ and ${\tilde{\Sigma}(0)
\times\!\!\!\!\!\!\supset Sp(32)}$ respectively. It is shown in
the Appendix that, precisely for $s=0$, both ${\tilde{\Sigma}(0)
\times\!\!\!\!\!\!\supset SO(1,10)}$ and ${\tilde{\Sigma}(0)
\times\!\!\!\!\!\!\supset Sp(32)}$ can be obtained from
$OSp(1|32)$ by the expansion method \cite{AIPV02}; they are given,
respectively, by the expansions $Osp(1|32)(2,3,2)$ and
$Osp(1|32)(2,3)$.

To trivialize the cocycle (\ref{a4=0}) over the $\tilde {\mathfrak
E}(s)$ enlarged superalgebra one considers the most general
ansatz\footnote{This was the starting point of \cite{D'A+F},
although for $\lambda=1$. Since more general possibilities --all
including an additional fermionic generator-- exist (cf.
\cite{BeSez95,JdA00}), one can motivate Eq.~(\ref{A3=Ans}) as
follows. As the $D=11$ superPoincar\'e algebra is not sufficient
to account for the gauge group structure of $D=11$ supergravity,
the next possibility would be to include the tensor charges
\cite{vHvP82,AGIT89} of the M-algebra. The ansatz would then be
Eq.~(\ref{A3=Ans}) for $\beta_1=\beta_2=\beta_3=0$ (no
$\eta^\alpha$), where only the first term may reproduce, under the
action of $d$, the bifermionic four form $a_4$, Eq. (\ref{a4=}).
This would fix $\lambda$ to be one. However, such an ansatz still
does not allow to obtain an $A_3$ obeying the FDA with
(\ref{R4=}). A new fermionic one-form $\eta^\alpha$ is thus
unavoidable and its inclusion provides a new contribution $\propto
\omega_4$, thus allowing for $\lambda\not=1$.} for the three-form
$A_3$ expressed in terms of wedge products of $e^a$,
$\psi^\alpha$; $B^{a_1a_2}$, $B^{a_1 \ldots a_5}$, $\eta^\alpha$,
\begin{eqnarray}
\label{A3=Ans} & \!\!\! 4A_3= \lambda B^{ab} \wedge e_a \wedge e_b
\; - \alpha_1 B_{ab} \wedge B^b{}_c \wedge B^{ca} - \alpha_2
B_{b_1a_1\ldots a_4} \wedge B^{b_1}{}_{b_2} \wedge B^{b_2a_1\ldots
a_4}  \hspace{1.8cm} \nonumber \\
&- \alpha_3  \epsilon_{a_1\ldots a_5b_1\ldots b_5c} B^{a_1\ldots
a_5} \wedge B^{b_1\ldots b_5} \wedge e^c -
 \alpha_4
\epsilon_{a_1\ldots a_6b_1\ldots b_5} B^{a_1a_2 a_3}{}_{c_1c_2}
\wedge B^{a_4a_5a_6c_1c_2} \wedge B^{b_1\ldots b_5}
\nonumber \\
&- 2i \psi^\beta \wedge \eta^\alpha \wedge \left( \beta_1 \, e^a
\Gamma_{a\alpha\beta} -i  \beta_2 \, B^{ab} \Gamma_{ab\;
\alpha\beta} + \beta_3 \, B^{a_1\ldots a_5} \Gamma_{a_1\ldots
a_5\; \alpha\beta} \right) \; ,
\end{eqnarray}
and looks for the values of the constants $\alpha_1, \ldots,
\alpha_4$, $\beta_1, \ldots, \beta_3$ {\it and} $\lambda$ such
that $dA_3=a_4$ in Eq.~(\ref{a4=}) provided $e^a$, $\psi^\alpha$,
$B^{a_1a_2}$, $B^{a_1\ldots a_5}$ and $\eta^\alpha$ are MC forms
obeying   (\ref{MCTa=}), (\ref{dB2=})--(\ref{deta=}) (we do not
distinguish notationally in Eq.~(\ref{A3=Ans}) and below between
the MC one-forms and the one-form gauge fields, nor between $A_3$
and $\tilde{\omega}_3$). If a solution exists, then
Eq.~(\ref{A3=Ans}) for the appropriate values of the constants
$\alpha_1, \ldots, \beta_3$ and $\lambda$ also provides an
expression for a composite $A_3$ satisfying (\ref{R4=}) in terms
of the one-forms obeying the FDA Eqs.~(\ref{Ta=}), (\ref{Tal=}),
(\ref{Rab=}), (\ref{cBab})--(\ref{cBal}). This is so because given
a Lie algebra through its MC equations, the Jacobi identities also
guarantee that the algebra obtained by adding non-zero curvatures
is a gauge FDA.

The condition that (\ref{A3=Ans}) satisfies (\ref{a4=}) produces a
set of equations for the constants $\alpha_1,\,\ldots,\, \beta_3$
and $\lambda$ including $\delta$, $\gamma_1$ and $\gamma_2$ as
parameters\footnote{This system of eight equations $\beta_1 +
10\beta_2 - 6! \beta_3=0$, $\lambda - 2\delta\beta_1=1$, $\lambda-
2\gamma_1 \beta_1 - 2\delta\beta_2=0$,  $3\alpha_1 + 8
\gamma_1\beta_2=0$, $\alpha_2 - 10\gamma_1 \beta_3 - 10\gamma_2
\beta_2=0$, $\alpha_3 - \delta \beta_3 - \gamma_2 \beta_1=0$,
$\alpha_2 - 5!\, 10\gamma_2 \beta_3=0$, $\alpha_3 - 2\gamma_2
\beta_3=0$, $3 \alpha_4 + 10\gamma_2 \beta_3=0$, is essentially
that of \cite{D'A+F} once $\lambda$ is set equal to one.}. This
system has a nontrivial solution for
\begin{equation} \label{det}
\Delta=(2 \gamma_1-  \delta)^2 = 4s^2 \gamma_1^2 \not=0 .
\end{equation}
The general solution has the form
\begin{eqnarray}\label{sEq=g}
& \lambda =  {1 \over 5} \; \frac{s^2+2s+6}{s^2} \; , \;
\beta_1=-\frac{1}{10\gamma_1} \; \frac{2s-3}{s^2} \; , \;
\beta_2=\frac{1}{20\gamma_1} \; \frac{s+3}{s^2} \; , \;
\beta_3=\frac{3}{10 \cdot 6! \gamma_1} \; \frac{s+6}{s^2}  \; ,
\nonumber \\
& \alpha_1=-\frac{1}{15} \; \frac{2s+6}{s^2} \; ,  \;
\alpha_2=\frac{1}{6!} \; \frac{(s+6)^2}{s^2} \; , \;
\alpha_3=\frac{1}{5 \cdot 6! 5!} \; \frac{(s+6)^2}{s^2} \; , \;
\alpha_4=- \frac{1}{9 \cdot 6! 5!} \; \frac{(s+6)^2}{s^2} \; ,
\quad
\end{eqnarray}
and exists $\forall s \neq 0$, {\it i.e.}, for any $\delta$,
$\gamma_1$, $\gamma_2$ obeying (\ref{idg}) except, as mentioned
above, for $\delta= 2\gamma_1$, $\gamma_2 =2\gamma_1 /5!$ ($\Delta
=0$) which corresponds to $s=0$ in (\ref{s-def}). Thus, the
$\omega_4$ cocycle (\ref{a4=0}) can be trivialized
($\omega_4=d\tilde{\omega}_3$) over all the
$\tilde{\mathfrak{E}}(s)$ superalgebras when $s\not=0$; the
impossibility of doing it over $\tilde {\mathfrak E}(0)$ may be
related with the fact that just $\tilde {\mathfrak E}(0)$ has an
enhanced automorphism symmetry, $Sp(32)$. As a result, the
three-form field\footnote{One may show that the (abelian) gauge
transformation properties $\delta A_3=d \alpha_2$ can be
reproduced from the gauge transformation properties of the new
fields.} $A_3$ of the standard CJS $D=11$ supergravity can be
considered as a composite of the gauge fields of the
$\tilde{\Sigma}(s)$ supergroups, $s\not=0$. In this case, taking
the exterior derivatives of (\ref{A3=Ans}) with the constants in
(\ref{sEq=g}) one also finds the expression for $\mathbf{R}_4$ in
terms of the two-form FDA curvatures.

The two particular solutions in \cite{D'A+F} are recovered by
adjusting $s$ ({\it i.e.}, $\delta, \gamma_1$ in
Eq.~(\ref{s-def})) so that $\lambda=1$ in Eq. (\ref{sEq=g}). This
is achieved for $\delta =5\gamma_1$ ($\delta$ non vanishing but
otherwise arbitrary), or for $\delta=0$ (with $\gamma_1$ non
vanishing but otherwise arbitrary). Thus, the two D'Auria and
Fr\'e decompositions of $A_3$ are characterized by
\begin{eqnarray}\label{D'A+F1}
& \delta=5\gamma_1  \not=0 \; ,
 \;  \gamma_2={\gamma_1 \over 2\cdot 4!} \; , \qquad
(\Leftrightarrow \quad \tilde {\mathfrak E}(3/2)) \nonumber \\ &
\lambda=1 \; , \beta_1=0 \; , \beta_2={1\over 10\gamma_1 } \; ,
\beta_3={1\over 6! \, \gamma_1} \; ,  \alpha_1=- {4\over 15} \; ,
\alpha_2={25\over 6!}  \; ,  \alpha_3={1\over 6!\, 4! } \; ,
\alpha_4=-{1\over 54\, (4!)^2} \; , \quad
\end{eqnarray}
and
\begin{eqnarray}\label{D'A+F2}
&   \delta=0 \; , \; \gamma_1 \not=0 \; ,
 \;  \gamma_2={\gamma_1 \over 3\cdot 4!} \; ,
\qquad (\Leftrightarrow \quad \tilde {\mathfrak E}(-1)) \nonumber
\\ \!\!\!\!\!\!  & \lambda=1 , \beta_1={1\over 2\gamma_1 }  ,
\beta_2={1\over 10\gamma_1 }  ,  \beta_3={1\over 4^. 5! \,
\gamma_1}  , \alpha_1=- {4\over 15}  ,  \alpha_2={25\over 6!} \; ,
\alpha_3={1\over 6!\, 4! }  ,  \alpha_4=-{1\over 54\, (4!)^2}\; .
\;
\end{eqnarray}

It is worth noting that there is a specially simple trivialization
of $\omega_4$. It is achieved for the family element $\tilde
{\mathfrak E}(-6)$, characterized by $\gamma_2=0$,
\begin{eqnarray} \label{S-6}
\tilde {\mathfrak E}(-6) &:& \quad \delta \neq 0 \; , \qquad
\delta=-10 \gamma_1 \; , \;   \gamma_2=0 \; .
\end{eqnarray}
In $\tilde {\mathfrak E}(-6)$ the generator $Z_{a_1\ldots a_5}$ is
central (see Eq.~(\ref{[Z,Q]})) and does not play any r\^ole in
the trivialization of the $\omega_4$ cocycle. Indeed, for these
values of the parameters, Eqs.~(\ref{susyalg})--(\ref{idg}) allow
us to consider the $\tilde {\mathfrak E}_{min}$ superalgebra whose
extension by the central charge $Z_{a_1\ldots a_5}$ gives $\tilde
{\mathfrak E}(-6)$ in Eq.~(\ref{S-6}). It is the
$(66+64)$-dimensional superalgebra $\tilde {\mathfrak E}_{min}$,
\begin{eqnarray}\label{susymin}
\{Q_\alpha,Q_\beta\} &=& \Gamma^a_{\alpha\beta} P_a +
i\Gamma^{a_1a_2}_{\alpha\beta} Z_{a_1a_2} \; , \\
{} [ P_a , Q_\alpha ] &=&  -10 \gamma_1 \; \Gamma_{a \;
\alpha}{}^\beta Q^\prime_\beta \; , \qquad [ Z_{a_1a_2} ,
Q_\alpha]=i \gamma_1 \Gamma_{a_1a_2 \; \alpha}{}^\beta
Q^\prime_\beta\;,
\end{eqnarray}
associated with the most economic
$\tilde{\Sigma}_{min}=\Sigma^{(66|32+32)}$ extension of the
standard supertranslation group (rigid superspace) on which
$\omega_4$ becomes trivial. The values of Eq.~(\ref{S-6}) in
Eq.~(\ref{sEq=g}) give
\begin{eqnarray} \label{coeffminimal}
& \lambda=\frac{1}{6} \; ,  \beta_1={1\over 4! \gamma_1 } \; ,
\beta_2=-{1\over 2 \cdot 5! \gamma_1 } \; ,  \beta_3=0 \; ,
\alpha_1={1\over 90} \; ,  \alpha_2=0  \; , \alpha_3=0 \; ,
\alpha_4=0 \; , \quad
\end{eqnarray}
and one notices in Eq.~(\ref{A3=Ans}) that all the terms
containing $B^{a_1 \ldots a_5}$ are zero. This makes the
expression for $A_3$ simpler,
\begin{eqnarray}\label{A3minimal}
\label{A3=min} A_3 &=& {1\over 4!} B^{ab} \wedge e_a \wedge e_b \;
- {1\over 3^.5!} B_{ab} \wedge B^b{}_c \wedge B^{ca}
\nonumber \\
&- &
 {i \over 4^. 5! \, \gamma_1}  \psi^\beta \wedge \eta^\alpha
\wedge \left( 10 \, e^a \Gamma_{a\alpha\beta} + i \, B^{ab}
\Gamma_{ab\;\alpha\beta} \right) \;
\end{eqnarray}
and thus $\Sigma^{(66|32+32)}$ can be regarded as a minimal
underlying gauge supergroup of $D=11$ supergravity.

The other $s\not=0$ representatives of the
$\tilde{\mathfrak{E}}(s)$ family are similar, although not
isomorphic. For instance, the momentum generator is central for
$\tilde{\mathfrak{E}}(-1)$ while $Z_{ab}$ is central for
$\tilde{\mathfrak{E}}(\infty)\,(\gamma_1=0)$. They all trivialize
the $\omega_4$ CE cocycle and, hence, provide a composite
expression of $A_3$ in terms of one-form gauge fields of the
enlarged supergroup $\tilde{\Sigma}(s)$.

\section{Concluding remarks}

We have shown that the cocycle $\omega_4$ (Eq.~(\ref{a4=0})) on
the standard $D=11$ supersymmetry algebra $\mathfrak{E}^{(11|32)}$
may be trivialized on the one-parametric family of superalgebras
$\tilde{\mathfrak{E}}(s)$, for $s \neq 0$, defined by
Eqs.~(\ref{susyalg})-(\ref{idg}) or (\ref{Sigma(s)}). These
superalgebras are central extensions of the M-algebra (of
generators $P_a, Q_\alpha, Z_{ab}, Z_{a_1\ldots a_5}$) by a
fermionic charge $Q^\prime_\alpha$. Trivializing the
supertranslation algebra cohomology four-cocycle $\omega_4$ on
the larger superalgebra $\tilde{\mathfrak{E}}(s)$, so that
$\omega_4 = d\tilde{\omega}_3$, is tantamount to finding a
composite structure for the three-form field $A_3$ of the
standard Cremmer-Julia-Scherk supergravity \cite{CJS} in terms
of one-form gauge fields of $\tilde{\Sigma}(s)$, $A_3=A_3(e^a \,
, \, \psi^\alpha \; ; \; B^{a_1a_2}, B^{a_1 \ldots a_5} ,
\eta^\alpha \,)$, Eq.~(\ref{A3=Ans}) with (\ref{sEq=g}). Such an
expression is given by the same equation (\ref{A3=Ans}) that
describes the $\tilde{\omega}_3$ trivialization of the $\omega_4$
cocycle, in which the Maurer-Cartan forms of
$\tilde{\mathfrak{E}}(s)$ are replaced by one-forms obeying a
free differential algebra with curvatures, Eqs.
(\ref{Ta=})--(\ref{Rab=}), (\ref{cBab})--(\ref{cBal}). Thus one
may treat the standard CJS $D=11$ supergravity as a gauge theory
of the ${\tilde{\Sigma}(s) \times\!\!\!\!\!\!\supset SO(1,10)}$
supergroup for any $s\not=0$.

This fact was known before for two superalgebras \cite{D'A+F} that
correspond to $\tilde{\Sigma}(3/2)$, Eq.~(\ref{D'A+F1}), and
$\tilde{\Sigma}(-1)$, Eq. (\ref{D'A+F2}) (although the whole
family $\tilde{\mathfrak{E}}(s)$ that results from Eq.~(\ref{idg})
was defined in \cite{D'A+F}). In this respect the novelty of our
results is that, for $s\neq0$, any of the $\tilde{\Sigma}(s)$
supergroups may be equally treated as an underlying gauge
supergroup of the $D=11$ supergravity. A special representative of
the family of trivializations is given by
$\tilde{\mathfrak{E}}(-6)$ for which the $Z_{a_1\ldots a_5}$
generator is central. The expression for $A_3$ trivializing the
cocycle $\omega_4$ over $\tilde{\mathfrak{E}}(-6)$ is particularly
simple: it does not involve the one-form $B^{a_1\ldots a_5}$.
Thus, the smaller
$\tilde{\Sigma}_{min}=\tilde{\Sigma}^{(66|32+32)}$ may be
considered as the minimal underlying gauge supergroup of $D=11$
CJS supergravity.

All other representatives of the family $\tilde{\mathfrak{E}}(s)$
are equivalent, although they are not isomorphic. Their
significance might be related to the fact that the field
$B^{a_1\ldots a_5}$ is needed \cite{BPS04} for a coupling to BPS
preons, the hypothetical basic constituents of M-theory
\cite{BPS01}. In a more conventional perspective, one can notice
that the charges $Z_{ab}$ and $Z_{a_1\ldots a_5}$ can be treated
as topological charges \cite{AGIT89} of M2 and M5 branes. In the
standard CJS supergravity the M2-brane solution carries a charge
of the three-form gauge field $A_3$ thus it should have a
relation with the charge $Z_{ab}$; that is reflected by Eq.
(\ref{A3minimal}) for a composite $A_3$ field and especially by
its first term $B_{ab}\wedge e^a\wedge e^b$ given by the natural
three-form constructed from the $Z_{ab}$ gauge field $B^{ab}$.
Similarly, the $Z_{a_1\ldots a_5}$ gauge field $B^{a_1\ldots a_5}$
should be related to the six-form gauge field $A_6$ which is dual
to the $A_3$ field and is necessary to consider the action for the
coupling of supergravity to the M5 brane \cite{BBS}. One might
expect that this $A_6$ field could also be a composite of
one-forms with basic term (the counterpart of the first one in
Eq. (\ref{A3minimal})) of the form $B^{a_1\ldots a_5}\wedge
e_{a_1}\wedge \ldots \wedge e_{a_5}$. The r\^ole of the fermionic
central charge $Q^\prime_\alpha$ and its gauge field $\eta^\alpha$
in this perspective also requires further study. Notice that such
a fermionic central charge is also present in the Green algebra
\cite{Green} (see also \cite{Sie94,BeSez95,JdA00}).

Although the presence of a full family of superalgebras
$\tilde{\mathfrak{E}}(s)$ --rather than a unique one--
trivializing the standard $\mathfrak{E}^{(11|32)}$ algebra
four-cocycle $\omega_4$, suggests that the obtained underlying
gauge symmetries of $D=11$ supergravity may be incomplete (this is
almost certainly the case if one considers the symmetries of
M-theory), the singularity of the $\tilde{\mathfrak{E}}(0)$ case
looks a reasonable one. The $\tilde{\Sigma}(0)$ supergroup is
special because it possesses an enhanced automorphism symmetry
$Sp(32)$ and the full ${\tilde{\Sigma}(0)
\times\!\!\!\!\!\!\supset Sp(32)}$, that replaces the $D=11$
superPoincar\'e group, is the expansion $OSp(1|32)(2,3)$ of
$OSp(1|32)$ (Appendix). The other members of the
$\tilde{\Sigma}(s)$ family only have a $SO(1,10)$ automorphism
symmetry and are deformations of the $s=0$ element. Thus our
conclusion is that the underlying gauge group structure of $D=11$
supergravity is determined by a one-parametric nontrivial
deformation of ${\tilde{\Sigma}(0)\times\!\!\!\!\!\!\supset
SO(1,10)} \subset {\tilde{\Sigma}(0)\times\!\!\!\!\!\!\supset
Sp(32)}$.

We would like to conclude with two remarks. The first is that we
did not consider in the expression of the $A_3$ field (see
Eq.~(\ref{A3=Ans})) Chern-Simons--like contributions as
$B_{a_1a_2} \wedge {\cal B}_2^{a_1a_2}$, $B_{a_1 \ldots a_5}
\wedge {\cal B}_2^{a_1 \ldots a_5}$, etc. These clearly would not
affect our cocycle trivialization arguments; their presence would
modify the expression of the composite $\mathbf{R}_4$ by
topological densities (see \cite{HoWi96} and {\it
e.g.}~\cite{EvSa02}). The second is that, unlike the lower
dimensional versions, $D=11$ supergravity forbids a cosmological
term extension. The reason may be traced \cite{BaDeHeSe97} to a
cohomological obstruction due to the presence of the three-form
field $A_3$. It would be interesting to analyze the implications
of its composite structure for this problem. The application of
the results of the present letter, and in particular the
consequences of a composite structure of $A_3$ for $D=11$
supergravity and M-theory, will be considered elsewhere.

\bigskip

{\it Acknowledgments}. This work has been partially supported by
the research grants BFM2002-03681, BFM2002-02000 from the
Ministerio de Educaci\'on y Ciencia and from EU FEDER funds, by
the grant N 383 of the Ukrainian State Fund for Fundamental
Research, the INTAS Research Project N 2000-254, and by the Junta
de Castilla y Le\'on grant A085-02. M.P. and O.V. wish to thank
the Ministerio de Educaci\'on y Ciencia and the Generalitat
Valenciana, respectively, for their FPU and FPI research grants.
Discussions with D. Sorokin and M. Henneaux are also acknowledged.

\appendix

\section{}

\subsection{$\tilde{\Sigma}(0) {\times\!\!\!\!\!\!\supset} \,SO(1,10)$ as the
expansion $OSp(1|32)(2,3,2)$}

To apply the expansion method~\cite{H01,AIPV02}, it will be
sufficient here to consider the case in which the superalgebra
$\mathcal{G}$ admits the splitting $\mathcal{G}=V_0\oplus V_1
\oplus V_2$, where $V_0$, $V_2$ ($V_1$), are even (odd) subspaces
of dimension $\textrm{dim}\, V_p$, $p=0,1,2$,  and $V_0$ is a
subalgebra of $\mathcal{G}$. Then, a rescaling of the group
parameters $g^{i_p} \rightarrow \lambda^p g^{i_p}$, $i_p=1,\ldots,
\textrm{dim}\, V_p$, makes the MC forms $\omega^{i_p}(\lambda)$
corresponding to the $p$-th subspace $V_p$, with the natural
grading $\omega^{i_p}(-\lambda)=(-1)^p \omega^{i_p}(\lambda)$, to
expand as a series in $\lambda$ as
\begin{equation} \label{expWW}
\hspace{-1.5cm} \omega^{i_p}(\lambda)=\lambda^p \omega^{i_p , p} +
\lambda^{p+2} \omega^{i_p , p+2} + \lambda^{p+4} \omega^{i_p ,
p+4} + \ldots \qquad (p=0,1,2) \; .
\end{equation}
The insertion of these series into the MC equations of $\cal G$,
\begin{equation} \label{eq:MC}
d\omega^{i_p}=-\frac{1}{2} c_{j_q k_s}^{i_p} \, \omega^{j_q}
\wedge \omega^{k_s} \quad (p,q,s=0,1,2 \; ; \;
i_{p,q,s}=1,2,\ldots, \textrm{dim} \, V_{p,q,s}) \; ,
\end{equation}
produces a set of equations identifying equal powers in $\lambda$.
The equations involving only the $\omega^{i_p, \alpha_p}$ up to
certain orders $\alpha_p=N_p,\; p=0,1,2$
($\alpha_p=p,p+2\ldots,N_p$) will determine the MC equations of a
Lie algebra provided that the highest $\omega^{i_p, N_p}$ orders
retained satisfy
\begin{equation} \label{conWW}
N_0=N_1+1=N_2 \qquad \mathrm{or} \qquad N_0=N_1-1=N_2 \qquad
\mathrm{or} \qquad  N_0=N_1-1=N_2-2 \; .
\end{equation}
The dimension of this new Lie algebra, the {\it expansion}
$\mathcal{G}(N_0,N_1,N_2)$ of $\mathcal{G}$, is \cite{AIPV02}
\begin{equation} \label{eq:dimsuper}
\textrm{dim}\,
\mathcal{G}(N_0,N_1,N_2)=\left[(N_0+2)/{2}\right]
\textrm{dim}V_0 +
\left[(N_1+1)/{2}\right]\textrm{dim}V_1 +
\left[N_2/{2}\right]\textrm{dim}V_2 \quad .
\end{equation}

Consider now the MC equations of $\tilde{\mathfrak{E}}(0)$, Eqs.
(\ref{MCTa=}), (\ref{dB2=}), (\ref{dB5=}) and (\ref{MCSigma(s)})
for $s=0$,
\begin{eqnarray} \label{Sigma(s=0)}
d\eta^{\alpha} &=& -2\gamma_1 \psi^\beta \wedge \left( i \,
e^a \Gamma_{a\, \beta}{}^\alpha + \frac{1}{2} \, B^{ab}
\Gamma_{ab\, \beta}{}^\alpha + \frac{i}{5!} \,  B^{a_1\ldots a_5}
\Gamma_{a_1\ldots a_5 \beta}{}^\alpha \right) \; , \qquad
\end{eqnarray}
to which we might add the $\omega^{ab}$ terms that implement the
$SO(1,10)$ automorphisms. The superalgebra $osp(1|32)$ is defined
by the MC equations
\begin{eqnarray}
     d\rho^{\alpha\beta}=-i\rho^{\alpha\gamma}\wedge
\rho_\gamma{}^\beta-i\nu^\alpha\wedge \nu^\beta \quad , \qquad
d\nu^\alpha=-i\nu^\beta \wedge \rho_\beta{}^\alpha\; , \qquad
\alpha,\beta=1,\ldots,32 \quad , \label{MCosp132}
\end{eqnarray}
where $\rho^{\alpha\beta}$ are the $sp(32)$ bosonic one-forms
($\rho_\gamma{}^\beta=C_{\gamma \alpha} \rho^{\alpha \beta}$,
where $C_{\alpha \beta}$ is identified with the $D=11$
imaginary charge conjugation matrix) and
$\nu^\alpha$ are the fermionic ones. The decomposition
\begin{equation}
\rho^{\alpha\beta}=\frac1{32} \left(  \rho^a \Gamma_a -\frac{i}{2}
\rho^{ab} \Gamma_{ab}+ \frac{1}{5!} \rho^{a_1\dots a_5}
\Gamma_{a_1\dots a_5}\right)^{\alpha\beta} \; , \;\;
a,b=0,1,\ldots,10 \; ,\label{generalrho}
\end{equation}
is adapted to the splitting \cite{AIPV02} $osp(1|32)=V_0\oplus
V_1\oplus V_2$, where $V_0$ is generated by $\rho^{ab}$, $V_1$ by
$\nu^\alpha$ and $V_2$ by $\rho^a$ and $\rho^{a_1\dots a_5}$. The
series (\ref{expWW}) take here the form
\begin{eqnarray}
&&\nu^\alpha=\lambda\nu^{\alpha,1} +
\lambda^3\nu^{\alpha,3}+\cdots,\quad
\rho^{ab}=\rho^{ab,0}+\lambda^2 \rho^{ab,2}+\cdots ,\quad \rho^a
=\lambda^2 \rho^{a,2}+\cdots ,\nonumber\\ &&\rho^{a_1\dots
a_5}=\lambda^2 \rho^{a_1\dots a_5,2}+\cdots\;.  \label{fullMexp}
\end{eqnarray}
Choosing $N_0=2$, $N_1=3$, $N_2=2$ (in agreement with conditions
(\ref{conWW})) one obtains the MC equations of the expansion
$osp(1|32)(2,3,2)$:
\begin{eqnarray}   \label{ospmaurerd}
 && \hspace{-.6cm}     d\rho^{ab,0}=-\frac{1}{16}
 \rho^{ac,0}\wedge {\rho_c}^{b\,,0} \quad, \qquad
      d\rho^{a\,,2}=-\frac{1}{16} \rho^{b,2}\wedge
{\rho_b}^{a,0} -i \nu^{\alpha,1}
\wedge \nu^{\beta,1} \Gamma^{a}_{\alpha\beta} \quad, \nonumber\\
&&  \hspace{-.6cm}    d\rho^{ab,2}=-\frac{1}{16} \left(
\rho^{ac,0}\wedge {\rho_c}^{b,2} + \rho^{ac,2}\wedge
{\rho_c}^{b,0} \right) - \nu^{\alpha,1}
\wedge \nu^{\beta,1} \Gamma^{ab}_{\alpha\beta} \quad, \nonumber\\
&&  \hspace{-.6cm}    d\rho^{a_1\dots a_5}{}^{,2}=\frac{5}{16}
\rho^{b[a_1\dots a_4|\,,2} \wedge \rho_b{}^{|a_5],0}\nonumber - i
\nu^{\alpha,1} \wedge \nu^{\beta,1} \Gamma^{a_1\dots
a_5}_{\alpha\beta} \quad, \nonumber\\
&&  \hspace{-.6cm}    d\nu^{\alpha,1}=-\frac{1}{64}
{\nu}^{\beta,1} \wedge
\rho^{ab,0}{\Gamma_{ab}}_\beta{}^\alpha \quad, \nonumber\\
&& \hspace{-.6cm}     d\nu^{\alpha,3}=-\frac{1}{64}
{\nu}^{\beta,3} \wedge \rho^{ab,0}{\Gamma_{ab}}_\beta{}^\alpha  \nonumber \\ && \qquad -
\frac{1}{32} \nu^{\beta,1} \wedge {\left(i\rho^{a,2}\Gamma_a +
\frac{1}{2}\rho^{ab,2}\Gamma_{ab} + \frac{i}{5!}\rho^{a_1\dots
a_5,2} \Gamma_{a_1\dots a_5}\right)_\beta}^\alpha  \, .\qquad\quad
\end{eqnarray}
Setting $\rho^{ab,0}=-16 \omega^{ab}$, Eqs.~(\ref{ospmaurerd})
coincide with those of $\tilde{\mathfrak{E}}(0)
{+\!\!\!\!\!\!\supset} so(1,10)$ [see Eqs.~(\ref{MCTa=}),
(\ref{dB2=}), (\ref{dB5=}) and (\ref{Sigma(s=0)})], with the
further identifications $\rho^{a,2}=e^a$, $\rho^{ab,2}=B^{ab}$,
$\rho^{a_1 \cdots a_5,2}=B^{a_1\cdots a_5}$,
$\nu^{\alpha,1}=\psi^\alpha$ and
$\nu^{\alpha,3}=\eta^\alpha/64\gamma_1$ (notice that $\gamma_1
\neq 0$ just defines the scale of $Q^\prime_\alpha$). Thus, we
conclude that $\tilde{\Sigma}(0){\times\!\!\!\!\!\!\supset}
SO(1,10)\approx OSp(1|32)(2,3,2)$ of dimension
$2\cdot55+2\cdot32+473=647$ by Eq.~(\ref{eq:dimsuper}).

\subsection{ $\tilde{\Sigma}(0) {\times\!\!\!\!\!\!\supset} Sp(32)$ as the
expansion $OSp(1|32)(2,3)$}

Let $osp(1|32)=V_0\oplus V_1$ where $V_0$ ($V_1$) is generated by
$\rho^{\alpha\beta}$ ($\nu^\alpha$). Choosing $N_0=2$ and $N_1=3$
we obtain the expansion $osp(1|32)(2,3)$ defined by the MC
equations:
\begin{eqnarray}\label{MC-sp}
&& \hspace{-1cm}
d\rho^{\alpha\beta,0}=-i\rho^{\alpha\gamma,0}\wedge
\rho_\gamma{}^{\beta ,0} \quad, \quad
d\rho^{\alpha\beta,2}=-i
\left( \rho^{\alpha\gamma,0}\wedge \rho_\gamma{}^{\beta ,  2} +
\rho^{\alpha\gamma,2}\wedge \rho_\gamma{}^{\beta ,  0} \right)
-i\nu^{\alpha,1}\wedge \nu^{\beta,1} \;,
\nonumber\\
&& \hspace{-1cm} d\nu^{\alpha,1}=- i\nu^{\beta,1} \wedge
\rho_\beta{}^{\alpha , 0} \quad, \quad d\nu^{\alpha,3}=-i
\nu^{\beta,3} \wedge \rho_\beta{}^{\alpha ,  0} - i\nu^{\beta,1}
\wedge \rho_\beta{}^{\alpha,2}\;.
\end{eqnarray}
Identifying $\rho^{\alpha\beta,0}$ in (\ref{MC-sp}) with the
$sp(32)$ connection $\Omega^{\alpha\beta}$, Eqs.~(\ref{MC-sp}) are
those of ${\tilde{\mathfrak{E}}(0){+\!\!\!\!\!\!\supset} sp(32)}$
[see Eqs.~(\ref{compacts0})] with
$\rho^{\alpha\beta,2}={\cal E}^{\alpha\beta}$,
$\nu^{\alpha,1}=\psi^\alpha$ and
$\nu^{\alpha,3}=\eta^\alpha/64\gamma_1$. Further,
dim(${\tilde{\mathfrak{E}}(0) {+\!\!\!\!\!\!\supset}
sp(32)}$)$=528+64+528=\textrm{dim}\, osp(1|32)(2,3)$ by
Eq.~(\ref{eq:dimsuper}).

\end{document}